\documentclass[aps,prl,twocolumn,amsmath,showpacs,letterpaper,superscriptaddress]{revtex4}

\usepackage{graphicx,amsmath,amssymb,amsfonts,latexsym,color,dcolumn,bm}
\usepackage{verbatim}

\begin{document}

\newcommand{\beq}{\begin{equation}}
\newcommand{\eeq}{  \end{equation}}
\newcommand{\bea}{\begin{eqnarray}}
\newcommand{\eea}{  \end{eqnarray}}
\newcommand{\bit}{\begin{itemize}}
\newcommand{\eit}{  \end{itemize}}
\newcommand{\jmax}{j_{\text{max}}}
\newcommand{\lmax}{\lambda_{\text{max}}}

\providecommand{\abs}[1]{\left\lvert#1\right\rvert}
\providecommand{\norm}[1]{\lVert #1 \rVert}
\providecommand{\moy}[1]{\langle #1 \rangle}
\providecommand{\bra}[1]{\langle #1 \rvert}
\providecommand{\ket}[1]{\lvert #1 \rangle}
\providecommand{\braket}[2]{\langle #1 \rvert #2 \rangle}

%\title{Molecular orientation entanglement and temporal Bell-type inequalities} 
\title{Bell-type inequalities for cold heteronuclear molecules}

\author{P. Milman}
\affiliation{Laboratoire de Photophysique Mol\'{e}culaire du CNRS, Univ. Paris-Sud, B\^{a}timent 210, Campus d'Orsay, 91405 Orsay Cedex, France}
\affiliation{CERMICS, Ecole Nationale des Ponts et Chauss\'{e}es, 6 et 8 av. Blaise Pascale, Cit\'{e} Descartes, Champs-sur-Marne, 77455 Marne-la-Vall\'{e}e, France }

\author{A. Keller}
\affiliation{Laboratoire de Photophysique Mol\'{e}culaire du CNRS, Univ. Paris-Sud, B\^{a}timent 210, Campus d'Orsay, 91405 Orsay Cedex, France}

\author{E. Charron}
\affiliation{Laboratoire de Photophysique Mol\'{e}culaire du CNRS, Univ. Paris-Sud, B\^{a}timent 210, Campus d'Orsay, 91405 Orsay Cedex, France}

\author{O. Atabek}
\affiliation{Laboratoire de Photophysique Mol\'{e}culaire du CNRS, Univ. Paris-Sud, B\^{a}timent 210, Campus d'Orsay, 91405 Orsay Cedex, France}

\begin{abstract} 
We introduce Bell-type inequalities allowing for non-locality and entanglement tests with two cold heteronuclear molecules. The proposed inequalities are based on correlations between each molecule spatial orientation, an observable which can be experimentally measured with present day technology. Orientation measurements are performed on each subsystem at different times. These times play the role of the polarizer angles in Bell tests realized with photons. We discuss the experimental implementations of the proposed tests, which could also be adapted to other high dimensional quantum angular momenta systems.
\end{abstract}

\pacs{03.65.Ud;03.67.-a;33.20.Sn}

\maketitle

Entanglement and non-locality are two related peculiarities of quantum theory. Historical debates\,\cite{EPR} and the advent of quantum information theory, to which entanglement is an essential ingredient, increased the interest paid to those concepts considered to be among the main traits of quantum mechanics\,\cite{SCHRO}. The experimental realization of Bell\,\cite{BELL} and Bell-type inequality tests\,\cite{CHSH} is of extreme importance to confirm the validity of quantum theory in a loophole free context\,\cite{BELL_EXP}. Since entanglement and non-locality are related (non-locality implies entanglement but not all entangled states are non-local), such tests may equally serve as entanglement witnesses\,\cite{WIT}. The interest of such approach also lies in the fact that for arbitrary high dimensional systems one does not dispose of necessary and sufficient conditions for entanglement characterization. However, detecting quantum properties of high dimensional systems is of clear interest in atomic, molecular and optical physics.

A number of experimental proposals are aiming to throw some light to these questions by means of Bell-type tests performed in different continuous or multi-dimensional systems: it was shown recently that correlated atoms originated from the dissociation of a molecular Bose-Einstein condensate can be used to test the Einstein-Podolski-Rosen paradox as originally formulated\,\cite{REFEREE,BELL}. The dichotomization of measurement results performed in a continuous system can also lead to maximal violation of Bell-type inequalities in phase space\,\cite{PS}. Finally, discrete multi-dimensional systems also allow for non-locality tests\,\cite{HD}, and recent experiments show the violation of Bell-type inequalities with two effective \mbox{spin--$1$} particles\,\cite{HD_EXP}. In parallel, recent advances in single molecule manipulation and detection open the way to the controlled creation of molecular entanglement\,\cite{Hettich2002}. Moreover, cold polar molecules confined in optical or magnetic traps have been recognized as a promissing candidate for quantum information processing\,\cite{SICM}, especially when their rotational levels are used as qubits\,\cite{Charron2006,Yelin2006}. Rotational states are relatively long lived, allowing for short quantum gate implementation times: one can perform about $10^4$ gate operations before decoherence takes place. This excellent performance, when compared to cold collision based quantum gates\,\cite{COLLI}, originates from the strength of dipolar interactions.

In the present paper, we propose realistic non-locality and entanglement tests for a system composed of two cold and trapped heteronuclear diatomic molecules. As in usual Bell tests scenarios, measurements are performed independently on each molecule by observers placed far apart, so that no comunication between them is possible during the realization of the protocol. We show in the following that inequalities built from local realism assumptions can be violated by a set of entangled states using measurements of correlations between the two molecules spatial orientation. These inequalities can also be used as entanglement witnesses\,\cite{WIT} when the measurement conditions are loosen. 

The inequalities derived here follow the original formulation of Clauser, Horne, Shimony and Holt (CHSH)\,\cite{CHSH}. For a pair of spin--$1/2$ particles or equivalent two-level systems, it can be shown using local theories that
\begin{equation}\label{chsh}
|\langle \sigma_a \sigma_b\rangle+\langle \sigma_a \sigma_{b'}\rangle+\langle \sigma_{a'} \sigma_b\rangle-\langle \sigma_{a'} \sigma_{b'}\rangle|\leqslant 2,
\end{equation}
where $\sigma_{\alpha}$ is the Pauli matrix in the $\alpha$ direction. $a$ and $a'$ refer to the first particle while $b$ and $b'$ refer to the second one. In order to derive the bound appearing in Eq.\,(\ref{chsh}) one assumes that the measurements performed by each observer are independent and their results have a probability distribution which is a product of independent probabilities for each subsystem. Such probabilities can also depend on some random local variable. Details are discussed in many works, as\,\cite{GISIN} for instance. It can be shown that a collection of entangled states can violate Eq.\,(\ref{chsh}), and this experimental violation was observed using photon pairs entangled in polarization\,\cite{BELL_EXP}. In this case, the directions $a$, $a'$, $b$ and $b'$ refer to different orientations of polarizers placed before the detectors. The value of the maximal possible violation, $2\sqrt{2}$, was first derived by Cirel'son\,\cite{CIRELSON}.

%For the general case of a $N$--level system, the numerical value of the bound depends on the spectrum of the measured observable.

The Bell inequalities we propose here have a structure similar to Eq.\,(\ref{chsh}). They are based on molecular orientation correlation measurements instead of spin-like observables. We show that they are experimentally implementable using time delayed measurements.

{\it Molecular spatial orientation}\,: We suppose that the two linear molecules that compose our system behave like rigid rotors described by their angular properties at a given time $t$. An arbitrary two--molecule bipartite state $\ket{\psi_0}$ has been created at time $t=0$, after which it freely evolves. This state can be entangled using techniques as the ones proposed in\,\cite{Charron2006,Yelin2006}. No interaction between molecules is allowed for $t>0$. The Hamiltonian describing each individual molecule's free evolution is \mbox{$\hat{H}_i=\hat{J}_i^{2}/\hbar^{2}$,} where $i=1,2$. It is expressed in units of the rotational energy. $\hat{J}_i$ is the angular momentum operator and therefore the associated evolution operator is given by $\hat{U}_i(t)= e^{-i\pi \hat{H}_it/\hbar}$, where time $t$ is written in units of the rotational period $\tau$. $\hat{U}_i(t)$ is therefore time-periodic with period 1. The total molecular state is thus given by the wavefunction $\psi(\theta_1,\theta_2,\phi_1,\phi_2,t)\equiv \braket{\theta_1,\theta_2,\phi_1,\phi_2}{\psi(t)}$. $\theta_i$ and $\phi_i$ denote here the polar and azimuthal spherical coordinates of both molecules in the laboratory frame. For each subsystem of the molecular bipartite set, the orientation at time $t$ is defined as the average value $\bra{\psi_0} O_i(t) \ket{\psi_0}$ of the orientation operator $\hat O_i(t) = \hat{U}_2^{-1}(t) \otimes \hat{U}_1^{-1}(t) \cos(\theta_i)\, \hat{U}_1(t) \otimes \hat{U}_2(t)$. Note that here $\cos(\theta_i)$ is taken as an operator, in perfect analogy to the position operator and related functions. Orientation as defined above can be experimentally measured by recording each molecule's fragments angular distribution following a quasi-instantaneous molecular dissociation induced by ultrafast Coulomb explosion\,\cite{SAKAI}.

With an arbitrary accuracy, each molecule's state (subscripts will be omitted) can be considered to reside in a finite dimensional Hilbert space
$\mathcal{H}$ generated by the basis set $\{\ket{j,m};\; 0 \leqslant j \leqslant \jmax,\, \abs{m} \leqslant j\}$, where $\ket{j,m}$ are the eigenstates of $\hat{J}^2$ and $\hat{J}_{z}$. Since it is a priori possible to prepare entangled bipartite molecular states with $m=0$ using ultra-cold atomic photoassociation or Feshbach resonances followed by a dipolar interaction scheme\,\cite{SICM,Charron2006}, we have considered here, for simplicity, that the angular momentum projection $m$ is fixed at $0$ for each molecule. We have checked that our conclusions remain valid even if $m \neq 0$. The dimension of each molecule's Hilbert space $\mathcal{H}$ is then $\jmax +1$, and the states $\ket{j,m=0}$ will now simply be written as $\ket{j}$. The corresponding wavefunctions are the spherical harmonics $\braket{\theta}{j} = Y_{j0}(\theta)$. In the finite Hilbert space $\mathcal{H}$, the $\cos\theta$ operator is characterized by a discrete, non-degenerate spectrum of eigenvalues $\lambda_n$ ($0 \leqslant n \leqslant \jmax$), with corresponding eigenvectors $\ket{\lambda_n}$, also called {\it orientation eigenstates}. The two maximally oriented states $\ket{+}$ and $\ket{-}$ are the two eigenstates corresponding to the extreme eigenvalues $\pm \lmax$, where $\lmax = \text{Max} (\lambda_n)$. In the particular case of $\jmax = 1$, the maximally oriented states can be simply written in the basis of the angular momentum eigenstates $\ket{0}$ and $\ket{1}$ as $\ket{+} \propto \ket{0}+\ket{1}$ and $\ket{-} \propto \ket{0}-\ket{1}$. In this particular case, $\cos{\theta} \propto \hat{\sigma}_x$, and we can of course recover the well-know results of a two-level system.

We now describe how orientation measurements performed locally on each subsystem can be used for non-locality tests through inequalities analogous to
Eq.\,(\ref{chsh}).

{\it Temporal inequalities}\,: The orientation correlation between the two molecules is now defined as the average value of the two-particle operator $C(t_1,t_2) = O_1(t_1) \otimes O_2(t_2)$, measured at times $t_1$ and $t_2$. Note that the operator $\cos{\theta}$ is useful here for entanglement detection since it ``mixes'' different values of $j$ without affecting their projection $m$. Previous works have considered correlations between different values of the projection $m$ for a given (fixed) value of $j$ in other physical contexts\,\cite{HD,agarwal}.

The rotation of the molecular axis is described by the free evolution operators $\hat{U}_i(t)$. As coherent superpositions of a finite numbers of $\ket{j}$, the orientation eigenstates are not stationnary states of the free Hamiltonian. Time evolution therefore modifies these quantum superpositions, thus changing the molecular orientation, in exact analogy to the usual projection of the photon polarization with polarizers. We would like to stress here that other CHSH inequalities using free evolution instead of polarizers were studied in different contexts: in Ref.\,\cite{GISIN,KAONSGO}, they allow the detection of entanglement between products of decaying mesons. In Ref.\,\cite{LEGGET_TEMP_DISC}, they reveal quantum properties of single particles.

\begin{figure}[!t]
\centering
\includegraphics*[height=5cm,clip=true]{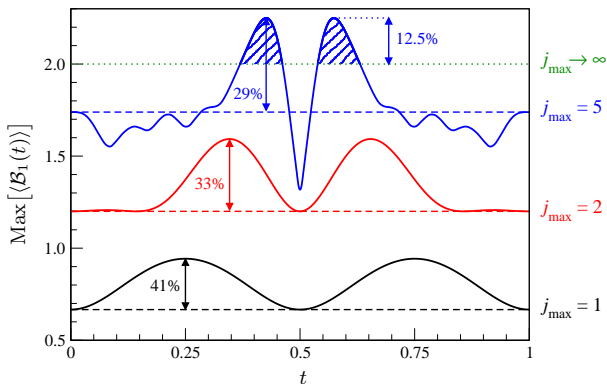}
\caption{(Color online) Maximal value of $\moy{{\cal B}_1(t)}$ in the subspace $\cal{H}$ with $j \leqslant \jmax$ as a function of $t$ in units of the rotational period $\tau$. The black, red and blue curves correspond to $\jmax=1$, 2 and 5 respectively. The associated separability thresholds are represented by dashed horizontal lines. The continuous case associated with the limit $\jmax \rightarrow +\infty$ is also shown as a dotted horizontal line.}
\label{fig1}
\end{figure}

By combining measurements realized at different times, one can define, in analogy to Eq.\,(\ref{chsh}), the operator
\begin{equation}
\label{cos}
{\cal B}_1(t) = C(t,t) + C(t,0) + C(0,t) - C(0,0).
\end{equation}
For simplicity, we have used a single time variable $t$ here in comparison with the four angles of Eq.\,(\ref{chsh}). We have verified that this limitation does not affect the generality of our results, since introducing four measurement times $t_i$ only increases the number of entangled states detected. In the framework of a local theory (LT), the operator ${\cal B}_1(t)$ obeys an inequality similar to Eq.\,(\ref{chsh}):
\beq
\label{eq:threshold}
\abs{\,\moy{{\cal B}_1(t)}_{\text{LT}}\,} \leqslant 2 \lmax^2\,, \quad \forall t > 0.
\eeq
Without loss of generality, we have also assumed that each particle state resides in the same finite dimensional Hilbert space $\mathcal{H}$. We note that Eq.\,(\ref{eq:threshold}) is valid for all possible values of $\jmax$, and that it can, in particular, be extended to the limit $\jmax \rightarrow + \infty$, in which case the spectrum of $\cos{\theta}$ forms a continuum. An interesting characteristic of the separability threshold\,(\ref{eq:threshold}) is its dependence on $\lmax$. In order to study the spectrum of ${\cal B}_1(t)$, we can numerically diagonalize this operator. We obtain, for each time $t$, its highest eigenvalue, which gives the maximal value of $\moy{{\cal B}_1(t)}$, and therefore the maximal violation of Eq.\,(\ref{eq:threshold}). This quantity depends on the dimensionality of the system, and we compare the amplitude of the violation for different values of $\jmax$ in Fig.\,\ref{fig1}. While $\jmax=1$ corresponds to a two-level system, $\jmax=2$ and $\jmax=5$ are two examples of a higher dimensional system where local realism can, in principle, be violated.

A first motivating result is that for a broad range of times, $\text{Max}[\moy{{\cal B}_1(t)}] > 2 \lmax^2\,$, thus violating Eq.\,(\ref{eq:threshold}). It proves that $\moy{{\cal B}_1(t)}$ is useful for non-locality tests. In particular, in the two dimensional case $\jmax=1$, $\text{Max}[\moy{{\cal B}_1(t)}]$ reaches its maximal value of $2\sqrt{2}\,\lmax^2$, in a similar fashion as for CHSH inequalities with two level systems. However, with increasing dimensionality, the maximum relative violation decreases. This effect is shown in Fig.\,\ref{fig1}: the maximal relative violation of Eq.(\ref{eq:threshold}) decreases from 41\% to 29\% when $\jmax$ varies from 1 to 5. In addition, the highest possible separability threshold given by Eq.\,(\ref{eq:threshold}) is obtained in a true infinite dimensional space, since in this case $\lmax \rightarrow 1$. The $\cos\theta$ eigenvalues then form a continuum, and Eq.\,(\ref{eq:threshold}) becomes a CHSH inequality for continuous measurement values. In a real experiment, one does sometimes not control the dimension of the subspace where entanglement is created, and the maximum threshold $2\lmax^2 \rightarrow 2$ should then be considered. Fig.\,\ref{fig1} shows that already for low values of $\jmax$ it is possible to violate this general threshold. Indeed, for $\jmax=5$, the maximum value of $\moy{{\cal B}_1(t)}$ is $9/4$, and for all values $\jmax \geqslant 5$, one can find entangled states violating Eq.\,(\ref{eq:threshold}). This result can be very useful in the realization of an experiment, and it is a consequence of two facts: high orientation (high values of $\lmax$) can be obtained in reduced angular momentum subspaces\,\cite{GRUPO} and entanglement enhances two particle orientation correlations\,\cite{Charron2006}.

\begin{figure}
\includegraphics*[height=5cm,clip=true]{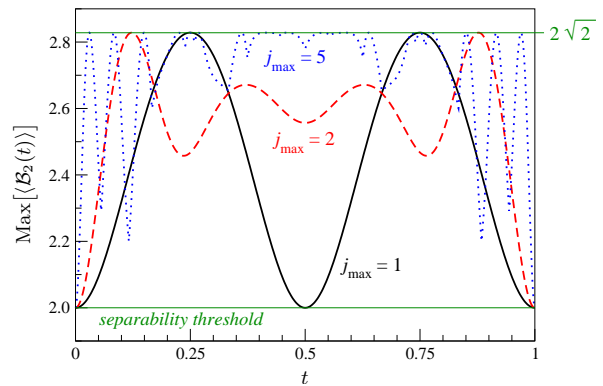}
\caption{(Color online) Maximal value of $\moy{{\cal B}_2(t)}$ as a function of $t$ in units of the rotational period $\tau$. The solid, dashed and dotted curves correspond to $\jmax=1$, 2 and 5 respectively. The separability and Cirel'son bounds are shown as horizontal lines.
\label{fig2}}
\end{figure}

We have shown that the operator defined by Eq.\,(\ref{cos}) allows not only for the realization of Bell-type tests in finite angular momentum
subspaces, but also when the size of the subspace is not a priori known. However, the maximal value of the violation decreases with dimensionality.
This may render the proposed test more difficult in very high dimensions. We can overcome this by a dichotomizing procedure, in which a high dimensional system is transformed into an effective two level one. The dichotomization is defined as follows: the states $\ket{\lambda_n}$ for which $\lambda_n = \bra{\lambda_n} \cos\theta\, \ket{\lambda_n} > 0$ are said to be positively oriented, while those for which $\lambda_n \leqslant 0$ are considered as negatively oriented. The orientation eigenstates are thus separated in two classes, $\ket{\lambda_{+}}$ and $\ket{\lambda_{-}}$, depending on the sign of their associated eigenvalue. We now define, for each molecule, the projectors on the subspaces of positive or negative orientation: $\Pi_{\pm} = \sum_{\lambda_{\pm}} \ket{\lambda_{\pm}} \bra{\lambda_{\pm}}$. The measured observable for the molecule $i$ is then $\Pi_i = \Pi_+ - \Pi_-$. This measurement consists in counting the asymmetry of the molecular angular distribution. For a given single-molecule state $\ket{\varphi} = \sum_{\lambda_+} c_{\lambda_+} \ket{\lambda_+} + \sum_{\lambda_-} c_{\lambda_-} \ket{\lambda_-}$, $\bra{\varphi} \Pi_i \ket{\varphi}$ can take any value in the interval $[+1,-1]$. The total two-molecule observable is defined as $\Pi = \Pi_1 \otimes \Pi_2$. We refer, as before, to two-molecule correlation measurements realized at two different times, using $\Pi(t_1,t_2) = \Pi_1(t_1) \otimes \Pi_2(t_2)$ where $\Pi_i(t_i) = \hat{U}_i^{-1}(t_i) \Pi_i\hat{U}_i(t_i)$. In analogy to Eq.\,(\ref{cos}), we now define the operator
\begin{equation}
\label{eq:B2}
{\cal B}_2(t) = \Pi(t,t) + \Pi(t,0) + \Pi(0,t) - \Pi(0,0).
\end{equation}
Since $\Pi(t_i,t_j)^2=1$, one can show that the highest value $\moy{{\cal B}_2(t)}$ can reach is given by the Cirel'son bound $2\sqrt{2}$\,\cite{CIRELSON}. In addition, in the framework of a local theory, we have
\beq
\label{dico} 
\abs{\moy{{\cal B}_2(t)}_{\text{LT}}} \leqslant 2\,, \quad \forall t > 0.
\eeq
Fig.\,\ref{fig2} shows the maximum value of $\moy{{\cal B}_2(t)}$ as a function of time for $\jmax=1$, 2 and 5. For $\jmax=1$ we obtain trivially the same result as with ${\cal B}_1$ (with a simple scaling factor). However, for $\jmax > 1$, measuring ${\cal B}_2$ has some advantages over ${\cal B}_1$. Cirel'son and locality bounds do not depend on dimensionality anymore, remaining valid even in the continuous case. Furthermore, Fig.\,\ref{fig1} and\,\ref{fig2} show that the range of time when ${\cal B}_2$ violates locality is much larger than the one of ${\cal B}_1$.

We now illustrate the principles of an ideal experiment leading to non-locality (and entanglement) tests using Eq.\,(\ref{eq:B2}). Our scenario consists of two molecules, created in a bipartite state $\ket{\psi_0}$ whose non-local properties one wishes to test. The molecules are spatially separated and evolve freely so that observers A and B can independently measure their orientation following the principles of\,\cite{SAKAI}. Measuring the average value of the dichotomized orientation\,(\ref{eq:B2}) requires only two detectors for each molecule, one placed in each hemisphere, determining whether molecules are positively or negatively oriented. Observers A and B perform such measurements at different times $t_i$. To ensure non-locality tests are realized, we impose that the measurements performed by each observer cannot causally affect the other's. For that, we suppose that A and B are separated by a distance $\ell$ such as $\ell > c\,\tau$, where $c$ is the speed of light and $\tau$ the molecule's rotational period. For typical values of $\tau$, this means that the observers should be separated by a distance greater than a few hundred $\mu$m. A possible experimental setup allowing for such molecular separation and manipulation are optical tweezers, currently used as individual atoms traps\,\cite{TWEEZERS}. The protocol is thus the following: after collecting orientation measurements for different times $t_i$, observers A and B compare their results and construct the quantity\,(\ref{eq:B2}). Using measurements lying in the interval determined by a rotational period $\tau$, they can experimentally infer the violation of\,(\ref{eq:B2}). In a simpler approach, we can also consider the situation where $\ell < c\,\tau$. In this case, correlations between measurements realized at different times and the proposed inequalities can serve as entanglement witnesses. This version of the protocol may find direct applications for molecules trapped in optical lattices\,\cite{MOLATT}.

In conclusion, we have introduced two experimentally realistic distinct high dimensional Bell-type inequalities based on the measurement of a quantity which is usually considered as being ``almost classical'': the molecular orientation. Both can be violated not only in the case of a restricted angular momentum subspace but also in the continuous limit. The proposed inequalities also present another original feature: orientation measurements are realized at different times for each particle. We have shown that dichotomization into two classes, namely positive and negative orientation, allows for a simple experimental measurement procedure that can be implemented with present day technologies using molecules in optical traps. This second approach also increases the number of detected entangled states. Our results open the perspective of entanglement detection and non-locality tests for high angular momentum systems in atomic and molecular physics.

\begin{acknowledgments}
During the realization of this work, P.M. was financially supported by the ACI ``{\it Simulations Mol\'{e}culaires"}.
This work was partially supported by the CEA, under contract No.\,LRC-DSM-0533.
\end{acknowledgments}

\end{document}